\documentclass[aps,twocolumn,showpacs,preprintnumbers,amsmath,amssymb,superscriptaddress]
{revtex4}
\usepackage{graphicx}
\usepackage{mathrsfs}
\usepackage{dcolumn}
\usepackage{bm}
\usepackage[bookmarks=false]{hyperref}

\begin{document}
\title
{Torque and conventional spin-Hall currents in two-dimensional
spin-orbit coupled systems: Universal relation and hyper-selection
rule}

\author{Tsung-Wei Chen}
\email{twchen@phys.ntu.edu.tw} \affiliation{Department of Physics
and Center for Theoretical Sciences, National Taiwan University,
Taipei 106, Taiwan}

\author{Guang-Yu Guo}
\email{gyguo@phys.ntu.edu.tw} \affiliation{Department of Physics
and Center for Theoretical Sciences, National Taiwan University,
Taipei 106, Taiwan}

\date{\today}

\begin{abstract}
We investigate torque and also conventionally defined spin-Hall
currents in two-dimensional (2D) spin-orbit coupled systems of
spin-$1/2$ particles within the linear response Kubo formalism. We
obtain some interesting relations between the conventional and
torque spin-Hall conductivities for the generic effective
Hamiltonian
$H_0=\epsilon_k^0+A(\mathbf{k})\sigma_x-B(\mathbf{k})\sigma_y$,
where
$A(\mathbf{k})=\eta^A_ik_i+\eta^A_{ij}k_ik_j+\eta^A_{ijl}k_ik_jk_l+\cdots$,
$B(\mathbf{k})=\eta^B_ik_i+\eta^B_{ij}k_ik_j+\eta^B_{ijl}k_ik_jk_l+\cdots$,
and $\eta$'s are the specific system-dependent coefficients.
Specifically, we find that in the intrinsic case the magnitude of
torque spin-Hall conductivity $\sigma^{\tau_z}_{xy}(0)$ is always
twice larger than the conventional spin-Hall conductivity
$\sigma^{s_z}_{xy}(0)$, and the two conductivities have the
opposite signs, i.e.,
$\sigma^{\tau_z}_{xy}(0)=-2\sigma^{s_z}_{xy}(0)$. This universal
relation, therefore, suggests that in the intrinsic case, the
total spin Hall conductivity $\sigma^z_{xy}(0)$ in the 2D systems
is equal to conventional spin Hall conductivity in magnitude but
has the opposite sign, namely,
$\sigma^z_{xy}(0)=\sigma^{\tau_z}_{xy}(0)+\sigma^{s_z}_{xy}(0)=
-\sigma^{s_z}_{xy}(0)$. This universal relation also holds in the
presence of an uniform in-plane magnetic field. We also find that
if the 2D systems are rotationally invariant, there exists a
hyper-angular momentum
$I_z=\left(\mathbf{k}\times\frac{\partial\theta}{\partial\mathbf{k}}\right)_zs_z+L_z$
which is conserved. Furthermore, the hyper-angular momentum
current $\langle\frac{1}{2}\{I_z,v_x\}\rangle$ vanishes, and this
leads to a hyper selection rule for the conventional spin-Hall
current. In particular, in the 2D k-linear Rashba and
wurtzite-type systems,  $I_z = s_z + L_z$, and the up(down)-spin
current would always be accompanied by the down(up)-orbital
angular momentum current (OAM). In the 2D k-cubic Rashba,
$I_z=3s_z+L_z$, and the hyper-selection rule is the same as in the
k-linear Rashba system. In the 2D k-linear Dresselhaus system, on
the other hand, $I_z = -s_z + L_z$, and the up(down)-spin current
would always be followed by the up(down)-OAM current.

\end{abstract}
 \maketitle
\section{Introduction}

Spin current generation is an important issue in the emerging
spintronics.\cite{pri98,Wolf01,Zutic04} Recent proposals of the
intrinsic spin Hall effect are therefore
remarkable~\cite{Mur03,Sinova04}. In the spin Hall effect (SHE), a
transverse spin current is generated in response to an electric
field in a system with spin-orbit coupling.~\cite{Dyako71,Hir90}
This effect has been considered to arise extrinsically, i.e., by
impurity scattering~\cite{Dyako71}. The scattering becomes
spin-dependent in the presence of spin-orbit coupling, and this
gives rise to the SHE. In the recent proposals, in contrast, the
SHE could arise intrinsically in hole-doped ($p$-type) bulk
semiconductors~\cite{Mur03} and also in electron-doped ($n$-type)
semiconductor heterostructures~\cite{Sinova04} due to intrinsic
spin-orbit coupling in the band structure. This intrinsic SHE
would thus provide a mechanism to generate electric driven spin
current without applied magnetic fields in semiconductors, which
can be more readily integrated with well-developed semiconductor
electronics. Recently, the spin accumulation at the edges of
semiconductor samples which is believed to be due to the SHE, has
been measured optically\cite{Kato04,Wund05,Chang07}. Further,
large SHE in metallic systems even at room temperature has been
detected electrically.\cite{Saito06,Valen06,Kimu07}

Many theoretical papers have been written addressing various
issues about the intrinsic SHE. In Ref. \onlinecite{Mur03}, the
SHE in the p-type GaAs semiconductor was explained as arising from
the k-space Berry curvature in response to the applied electric
field. This intrinsic SHE would lead to the possibility that the
spin-orbit coupling can be used to manipulate spin chirality in
semiconductors without dissipation. In Ref. \onlinecite{Mur04}, it
was shown that the SHE in the p-type GaAs semiconductor is robust
against the disorder based on the parity invariance of the
spherical Luttinger Hamiltonian. The Berry-phase-induced SHE was
also generalized to the case of spinning particles \cite{Bera06}.
In \cite{zha04}, an orbital-angular-momentum (OAM) Hall current is
predicted to exist in response to an electric field and is found
to cancel exactly the spin Hall current in the SHE. In
\cite{Guo05}, however, {\it ab inito} relativistic band structure
calculations show that the OAM Hall conductivity in p-type
semiconductors is one order of magnitude smaller than the spin
Hall conductivity, indicating no cancellation between the spin and
OAM Hall effects in bulk semiconductors.
The spin Hall conductivity in the two-dimensional (2D) k-linear
Rashba system has been shown to be suppressed by weak non-magnetic
disorder \cite{DisinR}. However, the spin-Hall conductivity
calculated with the consideration of the vertex correction due to
the impurity scattering, does not vanish, in general, and, e.g.,
in 2D k-cubic Rashba system \cite{Andre05}, 2D k-cubic wurtzite
system \cite{Mur04}, and 2D k-cubic Dresselhaus system
\cite{Malshu05}. Very recently, the large SHE in Pt metal at room
temperature~\cite{Kimu07} was also theoretically investigated and
was attributed to be an intrinsic one due to the band
anti-crossings near the Fermi level at the $L$ and $X$ symmetry
points in the Brillouin zone \cite{Guo08}.

The spin precession around the effective magnetic field caused by
spin-orbit coupling leads to the fundamental problem that the
conventionally defined intuitive spin current operator
$\frac{1}{2}\{\mathbf{v},s_z\}$ is not conserved. Therefore, how
to properly define the spin current operator has been intensively
studied in recent years \cite{Shi96,Zha05,Sun05,Vern07}. In view
of the spin continuity equation~\cite{Cul04}
$\frac{\partial\mathcal{S}_z}{\partial
t}+\nabla\cdot\mathbf{J}_s=\mathcal{T}_z$, Shi {\it et
al.}\cite{Shi96} recently provided a proper definition of
conserved spin current to resolve this issue. The effective
conserved spin current $\frac{d}{dt}(\mathbf{x}s_z)$ constructed
from the spin continuity equation is composed of two terms. One
term is the conventional intuitive spin current operator
$\frac{d\mathbf{x}}{dt}s_z$, and the other term
$\mathbf{x}\frac{ds_z}{dt}$ which is so-called torque spin current
comes from the spin precessional motion. Zhang {\it et
al.}\cite{Zha05} considered the spin Hall coefficients for three
widely studied semiconductor models, namely, 2D k-linear Rashba,
2D k-cubic Rashba and 3D Luttinger models, in the clean limit, and
found that the conserved spin Hall conductivities are dramatically
different from the conventional spin Hall conductivities. For
example, in the 2D systems, the conserved spin Hall conductivity
is equal to the conventional spin Hall conductivity in size but
has an opposite sign.\cite{Zha05} In Ref. \onlinecite{Sugi06}, the
results of calculations taking into account the conserved spin
current as well as impurity scattering effect for 2D k-linear
Rashba and k-cubic Rashba systems are reported. Recently, we
extended the conserved definition of spin current operator and
offered a proper definition of the OAM current
operator.\cite{Che06} We also found that in 2D Dresselhaus and
Rashba-Dresselhaus systems, the conserved spin Hall conductivity
is equal to the conventional spin Hall conductivity in size but
has an opposite sign.\cite{Che06}

Clearly, it is important to consider the new definition of spin
current~\cite{Shi96,Zha05,Sugi06,Che06} and it is of interest to
know the torque and hence conserved spin Hall coefficients in
other 2D systems.
In the present paper, therefore, we study the torque, conventional
and  conserved spin-Hall conductivities in all 2D spin-orbit
coupled systems described by a generic effective Hamiltonian [Eq.
(\ref{GenHam})] within the frequency-dependent Kubo linear
response theory. The generic effective Hamiltonian covers all
common 2D spin-orbit coupled systems used in the literature, such
as k-linear Rasha, Dresselhaus, Rashba-Dresselhaus, k-cubic
Dresselhaus and wurtzite-type Hamiltonians (Table I). We find two
interesting universal relations among the torque, conventional and
total conserved spin-Hall conductivities. Furthermore, we explore
possible connections between conventional spin current and orbital
motion of carriers and identify the existence of a conserved
hyper-angular momentum $I_z$ in rotationally invariant 2D
spin-orbit coupled systems. The conservation of the hyper-angular
momentum $I_z$ would lead to a hyper-selection rule which dictates
that the up(down)-spin state in the sense of
$(\mathbf{k}\times\frac{\partial\theta}{\partial\mathbf{k}})_zs_z$
would be accompanied by the down(up)-OAM state in these systems.

The present paper is organized as follows. In Sec. II we define a
generic effective Hamiltonian for 2D spin-orbit coupled systems
and calculate the time evolution of Pauli spin and position
operators in the Heisenberg picture. In Sec. III we calculate the
conventional and torque spin-Hall conductivities by using
frequency-dependent Kubo formulae and also present universal
relations between these conductivities. In Sec. IV we report our
finding that there exists a conserved hyper-angular momentum $I_z$
in the systems with the cylindrically symmetric energy dispersion.
We also demonstrate that the existence of $I_z$ leads to the hyper
selection rule for the conventional spin-Hall current. Our
conclusions are given in Sec. V. Three appendices to this paper
outline our derivation of the spin continuity equation, a proof of
conservation of $I_z$ and a proof of the vanishing of the $I_z$
current in 2D rotational invariant systems, respectively.


\section{Generic model Hamiltonian}
The effective Hamiltonian for spin-1/2 particles can be expressed
as a linear combination of Pauli matrices $\sigma_x$, $\sigma_y$
and $\sigma_z$. In 2D systems, we consider the following general
effective Hamiltonian,
\begin{equation}\label{GenHam}
H_0=\epsilon^{0}_{k}+A(\mathbf{k})\sigma_x-B(\mathbf{k})\sigma_y.
\end{equation}
where $\epsilon_k^0=\hbar^2k^2/2m$, is the single particle kinetic
energy, and functions $A(\mathbf{k})$ and $B(\mathbf{k})$ describe
the energy dispersion caused by spin-orbit interaction. In
general, $A(\mathbf{k})$ and $B(\mathbf{k})$ can be expressed as
$A(\mathbf{k})=\eta^A_ik_i+\eta^A_{ij}k_ik_j+\eta^A_{ijl}k_ik_jk_l+\cdots$
and
$B(\mathbf{k})=\eta^B_ik_i+\eta^B_{ij}k_ik_j+\eta^B_{ijl}k_ik_jk_l+\cdots$,
where $\eta$'s are the coefficients to be determined for each
specific system. The Einstein summation convention is used. The
general properties of coefficients $\eta$'s are determined by the
symmetry requirements. For instance, time reversal invariance of
spin current $\mathcal{J}=\frac{d(\mathbf{x}s_z)}{dt}$ requires
that $A(\mathbf{k})$ and $B(\mathbf{k})$ must be an odd function
of $\mathbf{k}$, i.e. $A(-\mathbf{k})=-A(\mathbf{k})$ and
$B(-\mathbf{k})=-B(\mathbf{k})$. This leads to the fact that the
spin dependent part of the Hamiltonian has no spatial inversion
symmetry. In appendix A, we show that the systems described by Eq.
(\ref{GenHam}) satisfy the spin continuity equation:
$\frac{\partial\mathcal{S}_z}{\partial
t}+\nabla\cdot\mathbf{J}_s=\mathcal{T}_z$. In Table \ref{UsuHam},
we list the specific functions $A(\mathbf{k})$ and $B(\mathbf{k})$
for several common 2D systems. However, we should stress here that
the following derivation is independent of the detailed forms of
$A(\mathbf{k})$ and $B(\mathbf{k})$.

We should emphasize that Eq. (1) is an effective Hamiltonian for
2D systems valid only near the Brillouin zone center, and is not a
bare Hamiltonian that describes the band structure of the whole
Brillouin zone. In other words, Eq. (1) is applicable to the 2D
semiconductor structures with the electron or hole pocket centered
at the Brillouin zone center such as p-type zinc-blende
semiconductors and n-type wurtzite nitrides, but not to the metals
with a complex Fermi surface such as platinum~\cite{Guo08}. In
writing the effective Hamiltonian Eq. (1), we made the assumption
that the particle spin-1/2 (or the pseudospin-1/2 for k-cubic
Rashba Hamiltonian) lies in the two dimensional plane. For these
spin-1/2 particles, we need only the two component Bloch wave
function and thus the effective Hamiltonian can be written as the
linear combination of Pauli matrices. Since the particle spin lies
in the plane, the spin splittings induced by bulk or structure
inversion asymmetry can be described by introducing the in-plane
components of the $k$-dependent effective magnetic field, in which
they are $A({\bf k})$ and $B({\bf k})$. The periodic potential and
spin-orbit coupling effect would enter the effective Hamiltonian
via the $A({\bf k})$ and $B({\bf k})$. The explicit forms of
$A({\bf k})$ and $B({\bf k})$ depend on the symmetries of the
underlying crystalline structure and band structure near the
Brillouin zone center. Since the 2D system we considered is time
reversal invariant (zero magnetic field), the spin-splitting would
result from the spatial inversion asymmetry (or structure
inversion asymmetry). This implies that $A(\mathbf{k})$ and
$B(\mathbf{k})$ are odd functions of k.

For the convenience of derivation, it turns out to be useful to
introduce a vector $\vec{M}=(M_x,M_y)$. The in-plane components of
vector $\vec{M}=(M_x,M_y)$ are
$M_x\equiv\cos\theta=\frac{B}{\Delta}$ and
$M_y\equiv\sin\theta=\frac{A}{\Delta}$. The Hamiltonian [Eq.
(\ref{GenHam})] can now be rewritten as
\begin{equation}\label{GenHam2}
H_0=\epsilon_k^0+\Delta(\vec{\sigma}\times\vec{M})_z,
\end{equation}
where $\Delta(\mathbf{k})=(A^2+B^2)^{1/2}$ is the energy
dispersion of spin-splitting determined by the explicit forms of
$A(\mathbf{k})$ and $B(\mathbf{k})$ (e.g., Table \ref{UsuHam}).
The vector Pauli matrix used in Eq. (\ref{GenHam2}) is
$\vec{\sigma}=(\sigma_x,\sigma_y)$. The eigenenergy of Eq.
(\ref{GenHam2}) is
$E_n(\mathbf{k})=\epsilon_k^0-n\Delta(\mathbf{k})$ and the
corresponding eigenvector is given by

\begin{equation}
    |n\mathbf{k}\rangle=\frac{1}{\sqrt{2}}\begin{pmatrix}
        e^{-i\theta(\mathbf{k})} \\ in
    \end{pmatrix},
    \label{eigenstate}
\end{equation}
where the $\theta(\mathbf{k})$ is
\begin{equation}\label{theta}
\theta(\mathbf{k})=\tan^{-1}\left(\frac{A(\mathbf{k})}{B(\mathbf{k})}\right)
\end{equation}
and the band index is denoted as $n=\pm$. It is straightforward to
show that $(\vec{\sigma}\times\vec{M})_z^2=(M_x^2+M_y^2)=1$. The
time evolution operator $exp(iH_0t/\hbar)$ can be further written
as
\begin{equation}
e^{iH_0t/\hbar}=e^{i\epsilon_k^0t/\hbar}\left[\cos\left(\frac{\Omega
t}{2}\right)+i(\vec{\sigma}\times\vec{M})_z\sin\left(\frac{\Omega
t}{2}\right)\right],
\end{equation}
where $\Omega=2\Delta/\hbar$. By using the definition of
Heisenberg picture for Schr\"{o}dinger operator $\mathcal{O}$,
$\mathcal{O}(t)=exp(iH_0t/\hbar)\mathcal{O}exp(-iH_0t/\hbar)$, one
can show that the time evolution of Pauli spin operators are given
by
\begin{equation}\label{Tpauli}
\begin{split}
\sigma_x(t)&=\sigma_x-M_x\sin(\Omega
t)\sigma_z+M_x(\vec{\sigma}\cdot\vec{M})[\cos(\Omega t)-1]\\
\sigma_y(t)&=\sigma_x-M_y\sin(\Omega
t)\sigma_z+M_y(\vec{\sigma}\cdot\vec{M})[\cos(\Omega t)-1]\\
\sigma_z(t)&=\cos(\Omega
t)\sigma_z+(\vec{\sigma}\cdot\vec{M})\sin(\Omega t).
\end{split}
\end{equation}
It can be shown that $\langle
n\mathbf{k}|(\vec{\sigma}\cdot\vec{M})|n\mathbf{k}\rangle=\langle
n\mathbf{k}|\sigma_z|n\mathbf{k}\rangle=0$ by the use of the
eigenstate in Eq. (\ref{eigenstate}). This means that the
expectation value of the z-component of the spin operator vanishes
in the absence of electric field. The time evolution position
operator can be written as
$\mathbf{x}(t)=\mathbf{x}(0)+\delta\mathbf{x}(t)$ and
\begin{equation}\label{Tposition}
\begin{split}
\delta\mathbf{x}(t)=&\left[\frac{\partial\epsilon_k^0}{\hbar\partial\mathbf{k}}+\frac{1}{2}(\vec{\sigma}\times\vec{M})_z\left(\frac{\partial\Omega}{\partial\mathbf{k}}\right)\right]t\\
&+\frac{1}{2}\frac{\partial\theta}{\partial\mathbf{k}}\left[(\cos(\Omega
t)-1)\sigma_z+(\vec{\sigma}\cdot\vec{M})\sin(\Omega t)\right],
\end{split}
\end{equation}
where $\mathbf{x}(0)$ is the initial condition and
$\delta\mathbf{x}(t=0)=0$. It can be shown that in the pure Rashba
system Eq. (\ref{Tposition}) would reproduce the result given in
Ref. \cite{Schl05}. The physical meaning of each term is as
follows. If the spin-orbit coupling vanishes, one has $\Delta=0$,
and the time evolution position operator Eq. (\ref{Tposition})
reduces to the free particle equation of motion
$\mathbf{x}(t)=\mathbf{x}(0)+\frac{\partial\epsilon_k^0}{\hbar\partial\mathbf{k}}t$.
The second term of Eq. (\ref{Tposition}) is the displacement
arising from the anomalous velocity in the presence of spin-orbit
coupling. The anomalous velocity plays an important role in the
anomalous Hall effect \cite{Crep01}. The third and fourth terms
have the oscillation behavior inducing the \emph{Zitterbewegung}
\cite{Schl05,Cser06,Bern07}.

\begin{table*}
\caption{Some common 2D systems where the effective Hamiltonian
can be described by Eq. (\ref{GenHam}). The $\Delta(\mathbf{k})$
describes the energy dispersion in the presence of spin-orbit
coupling, wherein
$\gamma(\phi)=\sqrt{\alpha^2+\beta^2-2\alpha\beta\sin(2\phi)}$,
$\kappa(\phi)=\frac{1}{2}\sin(2\phi)$, and
$\tan^{-1}\phi=\frac{k_y}{k_x}$. The $\sigma^{s_z}_{xy}(0)$ is the
conventionally defined spin-Hall conductivity. The pseudospin
angular momentum of the k-cubic Rashba hole system used in the
calculation of spin current is
$\mathbf{S}=\frac{3}{2}\hbar\vec{\sigma}$. Superscript * denotes
that the system is not rotationally invariant.}
\begin{ruledtabular}
\begin{tabular}{cccccc}
\label{UsuHam}
2-D system& $A(\mathbf{k})$ & $B(\mathbf{k})$ &  $\Delta(\mathbf{k})$& $\sigma^{s_z}_{xy}(0)$ &References \\
\hline \\Rashba & $\alpha k_y$  & $\alpha k_x$   & $\alpha k$ &
$\displaystyle\frac{-|e|}{8\pi}$ &\cite{byc84}
\\ \hline
\\Dresselhaus ([001]) & $\beta k_x$  & $\beta k_y$   & $\beta k$
&$\displaystyle\frac{|e|}{8\pi}$ &  \cite{dre55}
\\ \hline \\Dresselhaus ([110])$^*$ & $\rho k_x$  & $-\rho k_x$   & $\sqrt{2}\rho k\cos\phi$
&$0$ & \cite{Zutic04}
\\ \hline \\ Rashba-Dresselhaus$^*$   &  $\alpha k_y-\beta k_x$  & $\alpha k_x-\beta k_y$
& $k\gamma(\phi)$& $\displaystyle\frac{-|e|}{8\pi}
sign(\alpha^2-\beta^2)$ & \cite{Wong08,RandD,Che06}
\\ \hline \\k-cubic Rashba (hole)&$\displaystyle\frac{i\alpha_R}{2}(k_{-}^3-k_{+}^3)$&$\displaystyle\frac{\alpha_R}{2}(k_{-}^3+k_{+}^3)$& $\alpha_R k^3$&
$\displaystyle\frac{-9|e|\hbar^2}{16\pi^2m\alpha_R}(\frac{1}{k_F^+}-\frac{1}{k_F^-})$
&\cite{Loss05}
\\
\hline \\k-cubic Dresselhaus$^*$ & $\beta_D k_xk_y^2$ &$\beta_D
k_yk_x^2$& $\beta_D k^3\kappa(\phi)$& $\displaystyle\frac{|e|\hbar^2}{16\pi^2m\beta_D}\int\mathrm{d}\phi(\frac{\csc\phi}{k_F^+(\phi)}-\frac{\csc\phi}{k_F^-(\phi)})$&\cite{Malshu05} \\
\hline
\\Wurtzite type&$(\alpha_o+\beta_o k^2 )k_y$&
$(\alpha_o+\beta_o k^2)k_x$&$\alpha_o k+\beta_o k^3$&
$\displaystyle\frac{-|e|\hbar^2}{16m\pi}\frac{\tan^{-1}(\frac{\sqrt{\alpha_o\beta_o}(k_F^+-k_F^-)}{\alpha_o+\beta_ok_F^+k_F^-})}{\sqrt{\alpha_o\beta_o}}$&
\cite{Zor96,Chang07}
\end{tabular}
\end{ruledtabular}
\end{table*}

\section{Spin Hall conductivity}

As mentioned before, the conserved spin current is divided into
two terms:
\begin{equation}\label{Cspincur}
\frac{d}{dt}(\mathbf{x}s_z)=\frac{d\mathbf{x}}{dt}s_z+\mathbf{x}\frac{ds_z}{dt}.
\end{equation}
In addition to the conventional spin current
$\frac{d\mathbf{x}}{dt}s_z$, one have to introduce the torque spin
current $\mathbf{x}\frac{ds_z}{dt}$ in order to satisfy the
conserved spin continuity equation.  On the other hand, the time
reversal symmetry of the conserved spin current Eq.
(\ref{Cspincur}) would lead to spatial inversion asymmetry of spin
dependent part of Hamiltonian Eq. (\ref{GenHam}). This can be seen
as follows. From the commutator
$\frac{1}{i\hbar}[\sigma_z,H_0]=\Omega\vec{\sigma}\cdot\vec{M}$,
since the position operator is even under time reversal operation,
the invariance of torque spin current under time reversal symmetry
must require $A(-\mathbf{k})=-A(\mathbf{k})$ and
$B(-\mathbf{k})=-B(\mathbf{k})$. The time evolution of the
conserved spin current is
\begin{equation}
\begin{split}
\mathcal{J}(t)
&=\frac{1}{2}\{\mathbf{v}(t),s_z(t)\}+\frac{1}{2}\{\mathbf{x}(t),\frac{1}{i\hbar}[s_z,H_0](t)\}\\
&\equiv
\mathbf{J}^{s_z}(t)+\mathbf{J}^{\tau_z}(t)+\mathbf{J}^{\tau_0}(t),
\end{split}
\end{equation}
where
\begin{equation}\label{CSvec}
\mathbf{J}^{s_z}(t)=\frac{1}{2}\left\{\mathbf{v}(t),s_z(t)\right\}
\end{equation}
is the conventional spin current,
\begin{equation}\label{TSvec}
\mathbf{J}^{\tau_z}(t)=\frac{1}{2}\left\{\delta\mathbf{x}(t),\frac{1}{i\hbar}[s_z,H_0](t)\right\}
\end{equation}
is the torque spin current which is independent of the choice of
origin of coordinate system and
\begin{equation}\label{TSx}
\mathbf{J}^{\tau_0}(t)=\frac{1}{2}\left\{\mathbf{x}(0),\frac{1}{i\hbar}[s_z,H_0](t)\right\}
\end{equation}
is the other part of torque spin current which depends on the
initial choice of the origin of the coordinate system. The time
dependent part of the position operator $\delta\mathbf{x}(t)$ is
given by Eq. (\ref{Tposition}). It can be shown that $\langle
n\mathbf{k}|\mathbf{J}^{s_z}(t=0)|n\mathbf{k}\rangle=\langle
n\mathbf{k}|\mathbf{J}^{\tau_z}(t=0)|n\mathbf{k}\rangle=0$ and
$\langle
n\mathbf{k}|\mathbf{J}^{\tau_0}(t=0)|n\mathbf{k}\rangle=0$. This
leads to the fact that the conserved spin current vanishes at
$t=0$ as required, namely, $\langle
n\mathbf{k}|\mathcal{J}(t=0)|n\mathbf{k}\rangle=0$ in the absence
of external electric field. The value of Eq. (\ref{TSx}) is the
torque spin current with reference to the initial choice of the
origin of the coordinate system. We could choose the initial
position of the carrier as the origin of the coordinate system,
and as a result, Eq. (\ref{TSx}) would not contribute to the spin
accumulation. In that sense, the conserved spin current
$\mathcal{J}(t)$ can be divided into two terms
$\mathcal{J}(t)=\tilde{\mathcal{J}}(t)+\mathbf{J}^{\tau_0}$, where
\begin{equation}\label{TCSC}
\tilde{\mathcal{J}}(t)=\mathbf{J}^{s_z}(t)+\mathbf{J}^{\tau_z}(t)
\end{equation}
corresponds to the total spin current which is free from the
choice of the origin of the coordinate system. Eq. (\ref{TCSC})
could satisfy the initial condition, namely, $\langle
n\mathbf{k}|\tilde{\mathcal{J}}(t=0)|n\mathbf{k}\rangle=0$ because
it can be shown that $\langle
n\mathbf{k}|\mathbf{J}^{s_z}(t=0)|n\mathbf{k}\rangle=0$ and
$\langle
n\mathbf{k}|\mathbf{J}^{\tau_z}(t=0)|n\mathbf{k}\rangle=0$. The
frequency-dependent Kubo formula for a spatially homogeneous
electric field \cite{Schl04} is
\begin{equation}\label{Kubo}
\begin{split}
\sigma_{\mu\nu}(\omega)=&\frac{q/\hbar}{(\omega+i\eta)}\int^{\infty}_{0}\mathrm{d}te^{i(\omega+i\eta)t}\\
&\times\frac{1}{V}\sum_{n\mathbf{k}}f_{n\mathbf{k}}\langle
n\mathbf{k}|[J_{\mu}(t),v_{\nu}(0)]|n\mathbf{k}\rangle,
\end{split}
\end{equation}
where $q$ is the carrier charge, i.e., $q=-|e|$ for electrons, and
$f_{n\mathbf{k}}$ is the Fermi distribution at zero temperature.
The parameter $\eta$ is used to regularize the integral and the
direction of external electric field is denoted as index $\nu$. We
will calculate the conventional and torque spin-Hall
conductivities by using Eq. (\ref{Kubo}). We assume that the
electric field is applied in the y direction ($\nu\rightarrow y$).
The transverse spin current is composed of conventional and torque
spin-Hall currents,
$\tilde{\mathcal{J}}_x(t)=J^{s_z}_{x}(t)+J^{\tau_z}_x(t)$. The
conventional spin-Hall current in the x-direction is
$J^{s_z}_{x}(t)=\frac{1}{2}\{v_x,s_z\}(t)$, and it can be
evaluated as
\begin{equation}\label{CS}
J^{s_z}_{x}(t)=\frac{\hbar}{2}\tilde{v}_x\sigma_z(t),
\end{equation}
where $\tilde{v}_x$ is defined as $\tilde{v}_x=\hbar k_x/m$. The
torque spin-Hall current in the x direction is
$J^{\tau_z}_{x}(t)=\frac{\hbar}{2}\frac{1}{2}\{\delta
x(t),\frac{1}{i\hbar}[\sigma_z,H_0]\}$. After substitution of  the
commutator
$\frac{1}{i\hbar}[\sigma_z,H_0]=\Omega\vec{\sigma}\cdot\vec{M}$ to
the torque current and straightforward calculation, one can obtain
\begin{equation}\label{TS}
\begin{split}
J^{\tau_z}_{x}(t)&=\frac{\hbar}{2}\left[(\tilde{v}_xt)\Omega\vec{\sigma}(t)\cdot\vec{M}+\frac{\Omega}{2}\frac{\partial\theta}{\partial
k_x}\sin(\Omega t)\right],
\end{split}
\end{equation}
where $\delta x(t)$ given in Eq. (\ref{Tposition}) was used and
$\vec{\sigma}(t)=(\sigma_x(t),\sigma_y(t))$ wherein $\sigma_x(t)$
and $\sigma_y(t)$ are given in Eq. (\ref{Tpauli}). With the
definition of conserved spin current, the total spin Hall
conductivity is the sum of contributions of conventional and
torque spin-Hall currents,
\begin{equation}\label{TotalSC}
\sigma^{z}_{xy}(\tilde{\omega})=\sigma^{s_z}_{xy}(\tilde{\omega})+\sigma^{\tau_z}_{xy}(\tilde{\omega}),
\end{equation}
where $\tilde{\omega}=\omega+i\eta$. The first and second terms in
the right hand side of equality correspond to the conventional
spin Hall current and spin torque current, respectively. By using
Eq. (\ref{Kubo}) and
$v_y(0)=\tilde{v}_y+\frac{\partial\Delta}{\hbar\partial
k_y}(\vec{\sigma}\times\vec{M})_z+\frac{\partial\theta}{\hbar\partial
k_y}\Delta(\vec{\sigma}\cdot\vec{M})$, one can obtain

\begin{equation}\label{ConSC}
\sigma^{s_z}_{xy}(\tilde{\omega})=\frac{-q}{4\pi^2m}\int^{k_F^+}_{k_F^-}\mathrm{d}S_k\frac{\Delta(k_x\frac{\partial\theta}{\partial
k_y})}{\tilde{\omega}^2-\Omega^2},
\end{equation}
for the conventional spin-Hall conductivity and

\begin{equation}\label{TorSC}
\sigma^{\tau_z}_{xy}(\tilde{\omega})=\frac{-2q}{\pi^2m\hbar^2}\int^{k_F^+}_{k_F^-}\mathrm{d}S_k\frac{\Delta^3(k_x\frac{\partial\theta}{\partial
k_y})}{[\tilde{\omega}^2-\Omega^2]^2},
\end{equation}
for torque spin-Hall conductivity, where
$\tilde{\omega}=\omega+i\eta$, $\Omega=2\Delta/\hbar$, $k_F^{\pm}$
is Fermi momentum for band $n=\pm$ and
$\mathrm{d}S_k=k\mathrm{d}k\mathrm{d}\phi$. In the absence of
spin-orbit coupling, $\Delta=0$, the spin Hall conductivity
vanishes as one can see from Eq. (\ref{ConSC}) and Eq.
(\ref{TorSC}).

\subsection{The static limit: $\tilde{\omega}=0$}
In the intrinsic case and the static limit (i.e.
$\tilde{\omega}=0$), after some algebraic calculations, one can
obtain from Eq. (\ref{ConSC}) and Eq. (\ref{TorSC}) an exact
relation:

\begin{equation}\label{Result1}
\sigma^{\tau_z}_{xy}(0)=-2\sigma^{s_z}_{xy}(0).
\end{equation}
Thus, the torque spin-Hall conductivity is simply a constant (-2)
multiple of the conventional spin-Hall conductivity. This
universal relation implies that in the spin-1/2 2D systems with
spin-orbit coupling the magnitude of the torque spin-Hall
conductivity is always twice larger than the conventional
spin-Hall conductivity and has an opposite sign. The spin
z-component is not a constant of motion, as shown by the
commutator of the $s_z$ and $H_0$. We find that, by virtue of the
commutation properties of Pauli spin-1/2 matrices, $[s_z,H_0]$ is
a linear combination of the in-plane spin components, namely, x
and y components. We can rewrite the Hamiltonian as
$H_0=\epsilon_k^0+\vec{\sigma}\cdot\vec{\mathcal{B}}_{eff}$, where
$\vec{\mathcal{B}}_{eff}$ is the effective magnetic field in the
$\mathbf{k}$-space. The time derivative of spin z-component is
$d\sigma_z/dt=(-2)(\vec{\sigma}\times\vec{\mathcal{B}}_{eff})_z/\hbar$
and the quantity $-2$ on the right hand side of equality actually
yields the result that the magnitude of intrinsic torque spin-Hall
conductivity is always twice larger than conventional spin-Hall
conductivity. Physically, it is the spin precession that leads to
the result that the magnitude of torque spin Hall conductivity is
twice larger than conventional, and the sign of torque spin Hall
conductivity is opposite to conventional spin Hall conductivity.
We notice that the k-cubic Rashba system describing spin-3/2 heavy
hole also obeys Eq. (\ref{Result1}). In the k-cubic Rashba system
\cite{Loss05}, Pauli matrices operate on the states with spin-3/2
projection along the growth direction. In that sense, the k-cubic
Rashba system actually represents a pseudospin-1/2 system. We also
note that the spin-dependent part of the Hamiltonian of the
k-cubic Rashba is originally written as
$\frac{i\alpha_R}{2}(k^3_{-}\sigma_+-k^3_{+}\sigma_{-})$, where
$k_{\pm}=k_x\pm ik_y$ and $\sigma_{\pm}=\sigma_x\pm i\sigma_{y}$.
This can be rewritten as
$A(\mathbf{k})\sigma_x-B(\mathbf{k})\sigma_y$ with
$A=\frac{i\alpha_R}{2}(k_{-}^3-k_{+}^3)$ and
$B=\frac{\alpha_R}{2}(k_{-}^3+k_{+}^3)$. The simple relation
between intrinsic conventional and torque spin Hall conductivities
Eq. (\ref{Result1}) is independent of the detailed forms of spin
splitting (i.e. $A(\mathbf{k})$ and $B(\mathbf{k})$) and henece
the energy dispersion (i.e. $\Delta(\mathbf{k})$). The total
intrinsic spin Hall conductivity
$\sigma^{z}_{xy}(0)=\left[\sigma^{s_z}_{xy}(0)+\sigma^{\tau_z}_{xy}(0)\right]$
is then given by
\begin{equation}\label{Result2}
\sigma^{z}_{xy}(0)=-\sigma^{s_z}_{xy}(0).
\end{equation}
The sign of the total spin-Hall conductivity is always opposite to
the sign of the conventional spin-Hall conductivity. Even if
higher order k terms were included in the theoretical
calculations, the conclusions described above would still be true.
It should be emphasized that the validity of Eq. (\ref{Result2})
is independent of the sign of carrier charge. Interestingly, Eq.
(\ref{Result2}) suggests that the properties of the total
intrinsic spin-Hall conductivity can be characterized by the
conventional spin Hall conductivity only. Both Eq. (\ref{Result1})
and Eq. (\ref{Result2}) are the universal results for 2D
spin-orbit coupled systems. Let us now apply formulae Eq.
(\ref{ConSC}) and Eq. (\ref{TorSC}) to some specific systems. In
the Rashba-Dresselhaus system, for example, we can derive from
Eqs. (\ref{ConSC}) and (\ref{TorSC}) that
$\sigma^{s_z}_{xy}=(q/8\pi) sign(\alpha^2-\beta^2)$ and
$\sigma^{\tau_z}_{xy}=-(q/4\pi) sign(\alpha^2-\beta^2)$. The
results agree with previous works for the Rashba
system~\cite{Zha05} and for the Rashba-Dresselhaus
system~\cite{Che06}. We note that when $\alpha=\beta$, $A=-B$ and
$\theta$ is independent of $\mathbf{k}$. In this case, the
spin-Hall conductivity vanishes \cite{RandD,Che06}. For the
Dresselhaus Hamiltonian along the [110] direction (see Table. I.),
we have $A=-B$ and thus the spin-Hall conductivity also vanishes.
It has been shown that the [110] Dresselhaus Hamiltonian with
$\hat{e}_z$ along [001] direction is different from the
Rashba-Dresselhaus model with $\alpha=\beta$ only by a unitary
transformation \cite{Andri06}. The exact SU(2) spin rotation
symmetry has been investigated in both systems \cite{Andri06}.

We also note that Eq. (\ref{Result2}) is still true even if there
exists an uniform in-plane external magnetic field
$\vec{\mathcal{B}}$, as can be seen as follows. The interaction of
the spin and magnetic field is proportional to
$\vec{\sigma}\cdot\vec{\mathcal{B}}=\sigma_x\mathcal{B}_x-\sigma_y(-\mathcal{B}_y)$.
The Hamiltonian is now given by
$H=\epsilon^0_{k}+\sigma_x(A(k)+\mu_B\mathcal{B}_x)-\sigma_y(B(\mathbf{k})-\mu_B\mathcal{B}_y)$,
where $\mu_B$ is the Bohr magneton. We can redefine functions
$A'(\mathbf{k})$ and $B'(\mathbf{k})$ as
$A'(\mathbf{k})=A(k)+\mu\mathcal{B}_x$ and
$B'(\mathbf{k})=B(\mathbf{k})-\mu\mathcal{B}_y$, respectively.
Therefore, though the numerical values of Eq. (\ref{ConSC}) and
Eq. (\ref{TorSC}) may change, Eq. (\ref{Result1}) is still valid
in the presence of an uniform in-plane magnetic field.

Our predictions of, e.g., the conserved spin Hall current and
conductivity [Eq. (21)], can be tested by direct measurements of
the spin Hall current or conductivity. In particular, our
prediction that the total spin-Hall conductivity differs from the
conventional spin-Hall conductivity only in sign [Eq. (21)], could
be easily tested. As for the induction of a magnetic field by a
charge current, a spin-current would generate an electric
field~\cite{Sun04,Sun05}. Therefore, our prediction for the
conserved spin Hall current could be tested by comparison of the
measured spin current-induced electric field to the numerical
simulation for, e.g., a mesoscopic spin-orbit coupled system.
Another kind of experiments is to determine the inverse spin Hall
conductivity by measuring the charge current and Zeeman field
gradient~\cite{Shi96,Zha05}. The sign and magnitude of the
conserved spin Hall conductivity can then be obtained via the
Onsager relation~\cite{Shi96,Zha05}. Indeed, large spin Hall
effect in metallic systems at room temperature has recently been
detected by the method of inverse spin Hall
effect~\cite{Saito06,Valen06,Kimu07}. Hopefully, our interesting
predictions would stimulate measurements of inverse spin-Hall
effect in 2D semiconductor systems in the near future.



\subsection{Finite frequency case: $\tilde{\omega}\neq 0$}
At finite frequencies, the torque spin-Hall conductivity is not a
constant multiple of conventional spin-Hall conductivity.
Nonetheless, we find that Eq. (\ref{ConSC}) is related to Eq.
(\ref{TorSC}) by the following equation:
\begin{equation}\label{Result3}
\sigma^{\tau_z}_{xy}(\tilde{\omega})=(-2-\tilde{\omega}\frac{\partial}{\partial\tilde{\omega}})\sigma^{s_z}_{xy}(\tilde{\omega}).
\end{equation}
The second term in Eq. (\ref{Result3}) comes from the variation of
the $\mathbf{k}$-space effective magnetic field with the frequency
dependent external electric field. Taking into account the torque
spin current, we find that the total spin-Hall conductivity is
directly related to the conventional spin-Hall conductivity still.
Substituting Eq. (\ref{Result3}) into Eq. (\ref{TotalSC}), we
obtain the relationship between conventional spin-Hall
conductivity $\sigma^{s_z}_{xy}(\tilde{\omega})$ and total spin
Hall conductivity $\sigma^{z}_{xy}(\tilde{\omega})$ in the
presence of non-zero frequency-dependent electric field:
\begin{equation}\label{Result4}
\sigma^{z}_{xy}(\tilde{\omega})=-\frac{\partial}{\partial\tilde{\omega}}\left[\tilde{\omega}\sigma^{s_z}_{xy}(\tilde{\omega})\right].
\end{equation}
Eq. (\ref{Result4}) shows that the total spin-Hall conductivity
can be determined directly from the frequency spectrum of
conventional spin-Hall conductivity. Unlike the static limit, the
total ac spin-Hall conductivity is not proportional to the
conventional spin-Hall conductivity. It follows from Eq.
(\ref{Result4}) that the step function behavior of the
conventional spin-Hall conductivity would result in a large
response of the total spin-Hall conductivity. This large response
has recently been investigated in the Rashba-Dresselhaus system in
Ref. \onlinecite{Wong08}. It can be shown that Eqs. (\ref{ConSC})
and (\ref{TorSC}) for the Rashba-Dresselhaus system agree with the
results in Ref. \onlinecite{Wong08}. We also find that the simple
relation between the torque and conventional spin-Hall
conductivities [Eq. (\ref{Result3})] would be maintained even when
the external magnetic field is applied.

\section{constant of motion in rotationally invariant systems}
\subsection{Constant of motion}
If a 2D system is invariant under rotation about the z-axis, its
energy dispersion is cylindrically symmetric, i.e.,
$\Delta=\sum_{\ell}c_{\ell}k^{\ell}$ ($\ell=1,2,3,\cdots$). We
find that in this case, there exists a conserved quantity whose
operator $I_z$ is defined as
\begin{equation}\label{RealCC}
I_z=\left(\mathbf{k}\times\frac{\partial\theta}{\partial\mathbf{k}}\right)_zs_z+L_z.
\end{equation}
Let us call this quantity $I_z$ the hyper-angular momentum. We can
show that $I_z$ satisfies the following commutation relation (see
appendix B):
\begin{equation}\label{RealCC2}
[I_z,H_0]=0,
\end{equation}
where $s_z=\frac{\hbar}{2}\sigma_z$, $L_z=\hbar(xk_y-yk_x)$ is the
z-component of the OAM, and $H_0$ is given in Eq. (\ref{GenHam})
or Eq. (\ref{GenHam2}).
Interestingly, this implies that in the rotationally invariant
spin-orbit coupled systems, the flow of
$\left(\mathbf{k}\times\frac{\partial\theta}{\partial\mathbf{k}}\right)_zs_z$
would be accompanied by the orbital angular momentum $L_z$, and
the combination of these quantities is actually a constant of
motion. As a result, the spin current would in general be
accompanied by the OAM current because of the spin-orbit coupling.
In that sense, the external electric field would induce the
current of angular momentum
$(\mathbf{k}\times\frac{\partial\theta}{\partial\mathbf{k}})_zs_z$
and the OAM current simultaneously.

Further, it can be shown that the hyper-angular momentum Hall
current $\langle\frac{1}{2}\{I_z,v_x\}\rangle$ vanishes in the
steady state case within the linear-response Kubo formalism (see
appendix C). It follows that the up-spin (down-spin) state in the
sense of
$(\mathbf{k}\times\frac{\partial\theta}{\partial\mathbf{k}})_zs_z$
would be accompanied by the down-OAM (up-OAM) state, rendering the
hyper-angular momentum conserved. Let us now apply this result to
some specific systems. For the wurtzite type and k-linear Rashba
systems, $I_z=s_z+L_z$, i.e., the hyper-angular momentum is equal
to the total angular momentum. For the k-cubic Rashba hole system,
one would have $I_z=3s_z+L_z$. Therefore, Eq. (\ref{RealCC}) gives
the correct pseudospin angular momentum of hole which is
$\frac{3}{2}\hbar\sigma_z$. The conservation of hyper-angular
momentum in these systems would then lead to the result that the
spatial trajectory of a down-spin ($-\hat{e}_z$) carrier would
behave as having its orbital angular momentum pointed to
$+\hat{e}_z$ and vice versa (see Fig. 1. (a)). In other words, in
these systems, the situation of an up-spin (down-spin) state
accompanied by an up-OAM (down-OAM) trajectory is forbidden. This
is a hyper-selection rule that is present in the cylindrically
symmetric 2D spin-orbit coupled systems. It must be emphasized
that this hyper selection rule depends on the quantity
$(\mathbf{k}\times\frac{\partial\theta}{\partial\mathbf{k}})_z$.
For example, in the k-linear Dresselhaus system, we find that
$I_z=-s_z+L_z$. Therefore, in contrast to the k-linear and
wurtzite type systems, the hyper-selection rule in the k-linear
Dresselhaus system implies that an up-spin (down-spin) state would
be accompanied by an up-OAM (down-OAM) state [see Fig. 1. (b)].
The situation of an up-spin state accompanied by an down-OAM is
now forbidden in this Dresselhaus system. Nevertheless, we should
emphasize that the electric field-induced OAM current would not
result in magnetization accumulation at the edges of sample. This
is due to the fact that the OAM is not an intrinsic quantity of
electrons or holes, i.e., the magnetic moment associated with the
orbital angular momentum would vanish when the carrier velocity
reaches zero at the edges of sample, as can be understood from the
definition of $L_z$. Therefore, the magnetic moment accumulation
at the edges of sample would come from the spin angular momentum
only. In short, it is interesting to notice that the topological
quantity
$(\mathbf{k}\times\frac{\partial\theta}{\partial\mathbf{k}})_z$ is
a integer number for k-linear Rashba, k-linear Dresselhaus,
k-cubic Rashba and wurtzite-type system, whereas they are 1, -1,
3, 1, respectively. The topological number indicates that the
constant of motion in 2D rotationally invariant system is the
hyper-angular momentum rather than simply the total angular
momentum $s_z+L_z$.

It should be pointed out that although both hyper-angular momentum
conservation and hyper-selection rule exist in rotationally
invariant systems, this rotational symmetry may be broken when
higher order terms of $k_i$ are included in the $A(\mathbf{k})$
and $B(\mathbf{k})$ in Eq. (1). It would be necessary to include
the higher order terms of $k_i$ in Eq. (1) when the 2D
semiconductor systems considered have a very large carrier
concentration. When the weak symmetry-breaking higher order terms
do appear, the hyper-angular momentum [Eq. (24)] is no longer
conserved but the universal relation [Eq. (20)] still holds.

In the next subsection, we will describe the close relation
between Berry vector potential and hyper-angular momentum.

\begin{figure}
\begin{center}
\includegraphics[width=8cm,height=7cm]{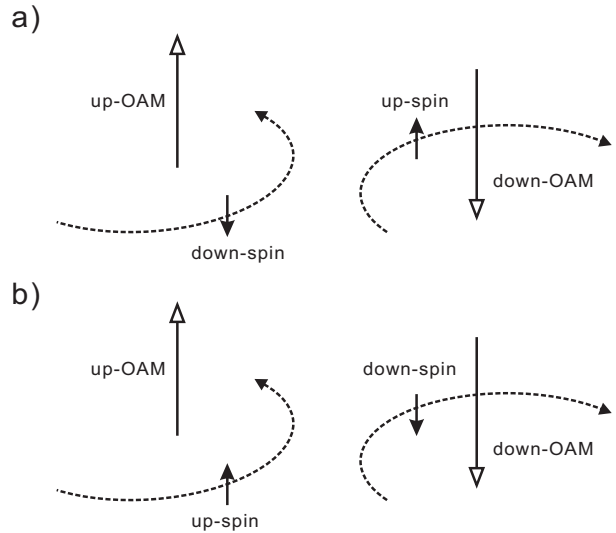}
\end{center}
\caption{
Schematic diagrams showing the relationship between spin and
orbital angular momentum (OAM) of a carrier. The direction of the
OAM is determined by the right-handed sense. The dotted line
illustrates the spatial trajectory of a carrier. (a) for the
Wurtzite type, k-linear and k-cubic Rashba systems and (b) for the
k-linear Dresselhaus system.}
\end{figure}

\subsection{Geometrical interpretation}
As mentioned above, the quantity
$(\mathbf{k}\times\frac{\partial\theta}{\partial\mathbf{k}})_zs_z$
together with the orbital angular momentum is conserved in a
rotationally invariant system. In the following, let us explain
that the quantity
$(\mathbf{k}\times\frac{\partial\theta}{\partial\mathbf{k}})_z$
actually comes from the topological properties of the Berry vector
potential. In the 2D systems with a cylindrically symmetric
dispersion,
the vector $\frac{\partial\theta}{\partial\mathbf{k}}$, 
in general, is perpendicular to the wave vector $\mathbf{k}$,
i.e.,
$\mathbf{k}\cdot\frac{\partial\theta}{\partial\mathbf{k}}=0$. For
example, it can be shown that the wurtzite type, k-linear Rashba,
k-cubic Rashba and k-linear Dresselhaus systems, the dot product
of $\mathbf{k}$ and $\frac{\partial\theta}{\partial\mathbf{k}}$ is
zero. In that sense, the three vectors $\mathbf{k}$,
$\frac{\partial\theta}{\partial\mathbf{k}}$ and
$\mathbf{k}\times\frac{\partial\theta}{\partial\mathbf{k}}$ form
an orthogonal frame fixed on the carrier (see Fig. 2.).

\begin{figure}
\begin{center}
\includegraphics[width=7cm,height=5cm]{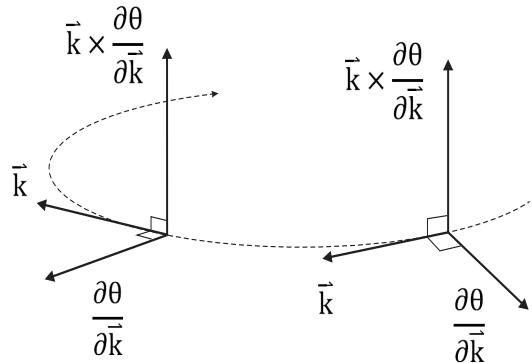}
\end{center}
\caption{A schematic diagram showing the relationship between
vectors $\mathbf{k}$,
$\mathbf{\frac{\partial\theta}{\partial\mathbf{k}}}$, and
$\mathbf{k}\times\mathbf{\frac{\partial\theta}{\partial\mathbf{k}}}$.}
\end{figure}

On the other hand, the Berry vector potential in a system
described by Hamiltonian Eq. (\ref{GenHam2}) can be written as
\begin{equation}
\mathscr{A}(\mathbf{k})=\langle
n\mathbf{k}|i\frac{\partial}{\partial\mathbf{k}}|n\mathbf{k}\rangle
=\frac{1}{2}\frac{\partial\theta}{\partial\mathbf{k}},
\end{equation}
where the eigenvector Eq. (\ref{eigenstate}) was used. The
projection of spin operator onto the in-plane axes contains two
terms. One term is the so-called spin helicity
$\vec{\sigma}\cdot\mathbf{k}$ in the 2D spin-orbit coupled system,
and the other is the projection of spin on the Berry vector
potential $\mathscr{A}\cdot\vec{\sigma}$. The quantity
$(\mathbf{k}\times\frac{\partial\theta}{\partial\mathbf{k}})_zs_z$
then comes from the noncommutativeness of the two in-plane
projections, viz,
\begin{equation}
\begin{split}
[\mathbf{k}\cdot\vec{\sigma},\mathscr{A}\cdot\vec{\sigma}]&=
[\mathbf{k}\cdot\vec{\sigma},\frac{1}{2}\frac{\partial\theta}{\partial\mathbf{k}}\cdot\vec{\sigma}]\\
&=i\left(\mathbf{k}\times\frac{\partial\theta}{\partial\mathbf{k}}\right)_z\sigma_z,
\end{split}
\end{equation}
where the commutation relations of Pauli matrices were used.  The
overall coefficient of
$(\mathbf{k}\times\frac{\partial\theta}{\partial\mathbf{k}})_zs_z$
cannot be determined by the commutation relation alone. However,
in the system with the cylindrically symmetric dispersion, the
hyper-angular momentum conservation forces the overall coefficient
of
$(\mathbf{k}\times\frac{\partial\theta}{\partial\mathbf{k}})_zs_z$
to be unity. Similar to the topological force induced by the
non-commutative position operator \cite{Mur03}, the spin part of
conservation of the hyper-angular momentum comes from the
non-commutative properties of $\vec{\sigma}\cdot\mathscr{A}$ and
$\vec{\sigma}\cdot\mathbf{k}$, whereas they are the projection of
spin on the two orthogonal axes. Finally, it must be stressed that
in the systems with the non-cylindrically symmetric dispersion,
the quantity
$(\mathbf{k}\times\frac{\partial\theta}{\partial\mathbf{k}})_zs_z$
can also be defined as the non-commutativeness of spin helicity
$\vec{\sigma}\cdot\mathbf{k}$ and $\vec{\sigma}\cdot\mathscr{A}$.
However, in that case, the quantity
$(\mathbf{k}\times\frac{\partial\theta}{\partial\mathbf{k}})_zs_z$
plus the orbital angular momentum is not conserved. Therefore, the
orbital motion of carrier does not accompany with
$(\mathbf{k}\times\frac{\partial\theta}{\partial\mathbf{k}})_zs_z$,
namely, the hyper-angular angular momentum is not conserved in
this case.

The conserved quantity in spin-orbit coupled systems can be
written as the sum of the spin and orbital terms. In the free
atomic case, the spin term is just the Pauli spin operator. When
the crystal environment is included, the spin term appears to be
different from the Pauli spin operator. It has a non-trivial
dependence on energy dispersion that arises from the spin-orbit
coupled effect. We find that the general coefficient is a
non-trivial multiplication of
$(\mathbf{k}\times\frac{\partial\theta}{\partial\mathbf{k}})_z$ in
rotationally invariant system. The quantity
$(\mathbf{k}\times\frac{\partial\theta}{\partial\mathbf{k}})_z$ is
proportional to the expectation value of the orbital angular
momentum.

It is of course not surprising that a conserved angular momentum
would exist in rotationally invariant systems. The conserved
angular momentum in spin-orbit coupled systems would be the sum of
the spin and orbital terms. In the free atom case, the spin term
is just the Pauli spin operator, and the sum of the spin and
orbital terms is indeed the total angular momentum. However, we
find here that in the presence of the crystal environment, the
spin term is not necessarily equal to the Pauli spin operator, but
has a non-trivial dependence on the energy dispersion [Eq. (24)]
that arises from the spin-orbit coupling effect. Therefore, we
would like to use the hyper-angular momentum here to differentiate
Eq. (24) from the well-known expression of the total angular
momentum in the free atom case. Furthermore, in the systems
described by the generic Hamiltonian Eq. (1), we find that the
coefficient in the spin term in Eq. (24) is related to the Berry
vector potential of the underlying band structure.

\section{Conclusions}
In conclusion, we have derived some interesting relations between
the conventional and torque spin-Hall conductivities for all 2D
spin-orbit coupled systems described by the generic effective
Hamiltonian Eq. (\ref{GenHam}) in the presence of
frequency-dependent electric field.
In particular, we find that an universal relation [Eq.
(\ref{Result2})] between total and conventional spin-Hall
conductivities, i.e., $\sigma^{z}_{xy}(0)=-\sigma^{s_z}_{xy}(0)$
for $\tilde{\omega}=0$. Eq. (\ref{Result2}) is independent of the
detailed form of energy dispersion $\Delta(\mathbf{k})$ (i.e.
$A(\mathbf{k})$ and $B(\mathbf{k})$), and hence its validity is
not restricted to the systems listed in Table. \ref{UsuHam}.
We also found that in 2D rotationally invariant systems, a
conserved hyper-angular momentum $I_z$ exists and the
hyper-angular momentum current vanishes. This would result in a
hyper-selection rule that the up-spin (down-spin) state in the
sense of
$(\mathbf{k}\times\frac{\partial\theta}{\partial\mathbf{k}})_zs_z$
would be accompanied by the down-OAM (up-OAM) state. Finally, we
explained that the spin dependent part of $I_z$ comes from the
noncommutative property of spin helicity
$\vec{\sigma}\cdot\mathbf{k}$ and $\vec{\sigma}\cdot\mathscr{A}$.


\section*{ACKNOWLEDGMENTS}
The authors would like to thank S.-Q. Shen and M. C. Chang for
useful discussions. The authors gratefully acknowledge financial
support from the National Science Council and NCTS of Taiwan.

\appendix
\section{Spin continuity equation}
In this appendix, we show that the Hamiltonian 
\begin{equation}\label{A1}
H=\frac{\mathbf{p}^2}{2m}+H_{so}+V(\mathbf{x})
\end{equation}
would satisfy the spin continuity equation:
\begin{equation}\label{A2}
\frac{\partial\mathcal{S}_z}{\partial
t}+\nabla\cdot\mathbf{J}^s=\mathcal{T}_z,
\end{equation}
where $H_{so}=A(\mathbf{p})\sigma_x-B(\mathbf{p})\sigma_y$
describes the spin-orbit interaction,
$V(\mathbf{x})=-q\mathbf{E}\cdot\mathbf{x}$ is the potential
induced by a homogeneous electric field $\mathbf{E}$,
$\mathcal{S}_z=\Psi^{\dag}s_z\Psi$ is the spin density,
$\mathbf{J}^s=Re\left[\Psi^{\dag}\frac{1}{2}\{\mathbf{v},s_z\}\Psi\right]$
is the conventional spin current and
$\mathcal{T}_z=Re\left[\Psi^{\dag}\frac{1}{i\hbar}[s_z,H_0]\Psi\right]$
is the source term of spin current. The carrier velocity is
defined as $\mathbf{v}=\frac{\partial H}{\partial\mathbf{p}}$. The
real part (imaginary part) of $[\cdots]$ is denoted as
$Re[\cdots]$ ($Im[\cdots]$), and $s_z=\frac{\hbar}{2}\sigma_z$. By
using
Schrodinger equation with the two-component wave function $\Psi=\begin{pmatrix} \phi_{\uparrow} \\
\phi_{\downarrow}
\end{pmatrix}$:
\begin{equation}\label{A3}
i\hbar\frac{\partial\Psi}{\partial t}=H\Psi,
\end{equation}
one can obtain:
\begin{equation}\label{A4}
\begin{split}
i\hbar\frac{\partial}{\partial
t}(\Psi^{\dag}s_z\Psi)=&\left\{-(\frac{\mathbf{p}^2}{2m}\Psi)^{\dag}s_z\Psi+\Psi^{\dag}s_z(\frac{\mathbf{p}^2}{2m}\Psi)\right\}\\
&+\left\{-(H_{so}\Psi)^{\dag}s_z\Psi+\Psi^{\dag}s_zH_{so}\Psi\right\}\\
&+\left\{-(V(\mathbf{x})\Psi)^{\dag}s_z\Psi+\Psi^{\dag}s_zV(\mathbf{x})\Psi\right\}.
\end{split}
\end{equation}
We note that
$[(H_{so}+V(\mathbf{x}))\Psi]^{\dag}s_z\Psi=[\Psi^{\dag}s_z(H_{so}+V(\mathbf{x}))\Psi]^{\dag}$
because the Pauli spin matrix $\sigma_z$ satisfies
$\sigma_z^{\dag}=\sigma_z$,i.e., is Hermitian. On the other hand,
one can also show that
\begin{equation}\label{A5}
[\Psi^{\dag}s_z\mathbf{p}^2\Psi-(\mathbf{p}^2\Psi)^{\dag}s_z\Psi]=2\mathbf{p}\cdot
Re[\Psi^{\dag}\mathbf{p}s_z\Psi].
\end{equation}
After substituting Eq. (\ref{A5}) into Eq. (\ref{A4}), one gets
\begin{equation}\label{A6}
\begin{split}
i\hbar\frac{\partial}{\partial
t}(\Psi^{\dag}s_z\Psi)=&\mathbf{p}\cdot
Re[\Psi^{\dag}\frac{\mathbf{p}}{m}s_z\Psi]\\
&+2iIm[\Psi^{\dag}s_zH_{so}\Psi]+2iIm[\Psi^{\dag}s_zV(\mathbf{x})\Psi].
\end{split}
\end{equation}
We also note that $s_zH_{so}$ can be written as
$s_zH_{so}=\frac{1}{2}\{s_z,H_{so}\}+\frac{1}{2}[s_z,H_{so}]$,
where the first term vanishes because the Pauli matrices satisfy
$\{\sigma_i,\sigma_j\}=2\delta_{ij}$. One can obtain
\begin{equation}\label{A7}
\begin{split}
Im[\Psi^{\dag}s_zH_{so}\Psi]&=Im\left[\Psi^{\dag}\frac{1}{2}[s_z,H_{so}]\Psi\right]\\
&=\frac{\hbar}{2}Re\left[\Psi^{\dag}\frac{1}{i\hbar}[s_z,H_{so}]\Psi\right].
\end{split}
\end{equation}
The last term of Eq. (\ref{A6}) vanishes because $V(\mathbf{x})$
is real as required by the Hermitian property of Hamiltonian Eq.
(\ref{A1}).
Finally, substitution of Eq. (\ref{A7}) into Eq. (\ref{A6}) yields
\begin{equation}\label{A8}
\begin{split}
\frac{\partial}{\partial t}(\Psi^{\dag}s_z\Psi)=&-\nabla\cdot
Re\left[\Psi^{\dag}\frac{1}{2}\{\mathbf{v},s_z\}\Psi\right]\\
&+Re\left[\Psi^{\dag}\frac{1}{i\hbar}[s_z,H_{so}]\Psi\right],
\end{split}
\end{equation}
where the commutation relation $\{\frac{\partial
H_{so}}{\partial\mathbf{p}},s_z\}=0$ was used. Eq. (\ref{A8}) is
the desired spin continuity equation. The average spin torque
vanishes \cite{Zha05}, and hence we have
$\int\mathrm{d}\mathbf{x}\mathcal{T}_z=0$. The spin torque density
can be written as the divergence of spin torque dipole density
$\mathbf{P}_{\tau}(\mathbf{x})$, namely,
$\mathcal{T}_z=-\nabla\cdot\mathbf{P}_{\tau}(\mathbf{x})$. On the
other hand, the spin dipole density vanishes outside the sample,
and we have
$\int_{V}\mathrm{d}\mathbf{x}\mathbf{P}_{\tau}=\int_{V}\mathrm{d}\mathbf{x}(-\mathbf{x}\nabla\cdot\mathbf{P}_{\tau})=\int_{V}\mathrm{d}\mathbf{x}(\mathbf{x}\mathcal{T}_z)$.
Therefore, the spin dipole density can be written as
$\mathbf{P}_{\tau}(\mathbf{x})=Re[\Psi^{\dag}\frac{1}{2}\{\mathbf{x},\frac{ds_z}{dt}\}\Psi]$.
Finally, the effective conserved spin continuity equation can be
written as
\begin{equation}
\frac{\partial\mathcal{S}_z}{\partial
t}+\nabla\cdot\mathcal{J}(\mathbf{x})=0,
\end{equation}
where
$\mathcal{J}(\mathbf{x})=Re[\Psi^{\dag}\hat{\mathcal{J}}\Psi]$ and
the effective conserved spin current operator is
$\hat{\mathcal{J}}=\frac{1}{2}\{\mathbf{v},s_z\}+\frac{1}{2}\{\mathbf{x},\frac{ds_z}{dt}\}$
which is the sum of conventional and torque spin currents.

\section{Conservation of $I_z$}
In this appendix, we demonstrate that the hyper-angular momentum
$I_z$ defined in Eq. (\ref{RealCC}) is a conserved quantity when
the energy dispersion $\Delta$ is rotationally invariant. First,
it can be shown that the velocity operator can be written as
\begin{equation}\label{velo}
v_x=\frac{1}{i\hbar}[x,H_0]=\tilde{v}_x+\frac{\partial\Delta}{\hbar\partial
k_x}(\vec{\sigma}\times\vec{M})_z+\frac{\partial\theta}{\hbar\partial
k_x}\Delta(\vec{\sigma}\cdot\vec{M}),
\end{equation}
where $H_0=\epsilon_k^0+\Delta(\vec{\sigma}\times\vec{M})_z$ was
used. The y-component of velocity can be obtained by replacing the
index $x$ by $y$. The z-component of orbital angular momentum
operator is defined as $L_z=\hbar(xk_y-yk_x)$. Using the velocity
operator, the commutator $[L_z,H_0]$ is straightforwardly
evaluated as follows:
\begin{equation}\label{Lz}
\begin{split}
\frac{1}{i\hbar}[L_z,H_0]&=\hbar(v_xk_y-v_yk_x)\\
&=\Delta\left(\frac{\partial\theta}{\partial
k_x}k_y-\frac{\partial\theta}{\partial
k_y}k_x\right)(\vec{\sigma}\cdot\vec{M})\\
&+\left(\frac{\partial\Delta}{\partial
k_x}k_y-\frac{\partial\Delta}{\partial
k_y}k_x\right)(\vec{\sigma}\times\vec{M})_z.
\end{split}
\end{equation}
We now define the operator $I_z$ as
\begin{equation}\label{Iz}
I_z=\left(\mathbf{k}\times\frac{\partial\theta}{\partial\mathbf{k}}\right)_zs_z+L_z,
\end{equation}
where $\theta=\tan^{-1}\left(\frac{A}{B}\right)$ and
$s_z=\frac{\hbar}{2}\sigma_z$ is the Pauli spin operator. Using
Eqs. (\ref{Lz}) and (\ref{Iz}), we obtain
\begin{equation}\label{IzH}
\frac{1}{i\hbar}[I_z,H_0]=\left(\frac{\partial\Delta}{\partial
k_x}k_y-\frac{\partial\Delta}{\partial
k_y}k_x\right)(\vec{\sigma}\times\vec{M})_z,
\end{equation}
where
$\frac{1}{i\hbar}[\sigma_z,H_0]=\frac{2\Delta}{\hbar}\vec{\sigma}\cdot\vec{M}$
was used. Eq. (\ref{IzH}) is the main result of this appendix, and
it means that, in general, the $I_z$ operator is not a conserved
quantity. The right hand side of Eq. (\ref{IzH}) explicitly
depends on the form of energy dispersion. It is interesting to
note that the spin term of hyper-angular momentum (Eq. (\ref{Iz}))
is not the Pauli matrices with the multiplication of $\hbar/2$,
but the multiplication of
\begin{equation}
\left(\mathbf{k}\times\frac{\partial\theta}{\partial\mathbf{k}}\right)_z
\end{equation}
which is further explained in Sec. IV. B.

Now consider a 2D system with the cylindrically symmetric energy
dispersion that can be written as the power series of magnitude of
$\mathbf{k}$ denoted as $k=|\mathbf{k}|$, namely,
$\Delta=\sum_{\ell}c_{\ell}k^{\ell}$. The right hand side of
equality in Eq. (\ref{IzH}) then becomes
\begin{equation}
\left(\frac{\partial\Delta}{\partial
k_x}k_y-\frac{\partial\Delta}{\partial
k_y}k_x\right)=\sum_{\ell}c_{\ell}\ell
k^{\ell-1}\left(\frac{k_x}{k}k_y-\frac{k_y}{k}k_x\right)=0.
\end{equation}
Therefore, the hyper-angular momentum $I_z$ is a conserved
quantity in the rotationally invariant 2D systems.

\section{Null hyper-angular momentum current}
In this appendix, we will show that the hyper-angular momentum
current $\frac{1}{2}\{I_z,\mathbf{v}\}$ vanishes in the linear
response regime. In the static case, the Kubo formula can be
written as
\begin{equation}
\begin{split}
\sigma_{\mu\nu}=&~q\hbar\sum_{n\neq
n'}\sum_{\mathbf{k}}\frac{f_{n\mathbf{k}}-f_{n'\mathbf{k}}}{(E_n(\mathbf{k})-E_{n'}(\mathbf{k}))^2}\\
&~\times Im\langle n\mathbf{k}|J_{\mu}|n'\mathbf{k}\rangle\langle
n'\mathbf{k}|v_{\nu}|n\mathbf{k}\rangle,
\end{split}
\end{equation}
where both eigenstate $|n\mathbf{k}\rangle$ and eigenenergy
$E_{n}(\mathbf{k})$ are given in Sec II. In the following, the
external electric field is assumed to be applied in the $y$
direction, and we calculate the conductivity $\sigma_{xy}$. First
of all, the hyper-angular momentum  current can be divided into
two terms $J_x^{I_z}=J^{\mathscr{S}_z}_x+J^{L_z}_x$. First term is
the hyper-spin current
\begin{equation}
J^{\mathscr{S}_z}_x=\frac{1}{2}\{\left(\mathbf{k}\times\frac{\partial\theta}{\partial\mathbf{k}}\right)_zs_z,v_x\}
\end{equation}
corresponding to the hyper spin-Hall conductivity
$\sigma^{\mathscr{S}_z}_{xy}$. The second term of $J^{I_z}_z$ is
the orbital current
\begin{equation}
J^{L_z}_x=\frac{1}{2}\{L_z,v_x\}
\end{equation}
corresponding to the orbital-Hall conductivity
$\sigma^{L_z}_{xy}$. As a result, the hyper angular momentum Hall
conductivity can be written as
\begin{equation}\label{hyperCon}
\sigma^{I_z}_{xy}=\sigma^{\mathscr{S}_z}_{xy}+\sigma^{L_z}_{xy}.
\end{equation}
We first calculate the orbital-Hall conductivity. By using the
velocity operator Eq. (\ref{velo}), we have
\begin{equation}\label{ImO}
\begin{split}
&Im\langle n\mathbf{k}|J^{L_z}_{x}|-n\mathbf{k}\rangle\langle
-n\mathbf{k}|v_{y}|n\mathbf{k}\rangle\\
&=n\frac{\hbar k_x}{2m}\Delta\frac{\partial\theta}{\partial
k_y}\left(\frac{\partial\theta}{\partial\mathbf{k}}\times\mathbf{k}\right)_z\\
&-\frac{1}{2\hbar}Im\{i\Delta\frac{\partial\theta}{\partial
k_y}(i\frac{\partial A}{\partial k_x}+\frac{\partial B}{\partial
k_x})e^{-i\theta}\}\left(\frac{\partial\theta}{\partial\mathbf{k}}\times\mathbf{k}\right)_z
\end{split}
\end{equation}
The second term of Eq. (\ref{ImO}) does not contribute to the
orbital-Hall conductivity $\sigma^{L_z}_{xy}$ because of the even
power of band index $n$. The orbital-Hall conductivity
$\sigma^{L_z}_{xy}$ with substitution of Eq. (\ref{ImO}) gives
\begin{equation}\label{hyperLcon}
\sigma^{L_z}_{xy}=\frac{q\hbar^2}{16\pi^2m}\int^{k^{+}_F}_{k^{-}_F}dS_k\frac{k_x\frac{\partial\theta}{\partial
k_y}}{\Delta}\left(\frac{\partial\theta}{\partial\mathbf{k}}\times\mathbf{k}\right)_z.
\end{equation}
We now consider the hyper-spin Hall current. Taking into account
the hyper-spin and the velocity operator Eq. (\ref{velo}), we have
\begin{equation}\label{ImS}
\begin{split}
&Im\langle
n\mathbf{k}|J^{\mathscr{S}_z}_{x}|-n\mathbf{k}\rangle\langle
-n\mathbf{k}|v_{y}|n\mathbf{k}\rangle\\
&=n\frac{\hbar k_x}{2m}\Delta\frac{\partial\theta}{\partial
k_y}\left(\mathbf{k}\times\frac{\partial\theta}{\partial\mathbf{k}}\right)_z.
\end{split}
\end{equation}
Inserting Eq. (\ref{ImS}) into $\sigma^{\mathscr{S}_z}_{xy}$, we
obtain
\begin{equation}\label{hyperScon}
\sigma^{\mathscr{S}_z}_{xy}=\frac{q\hbar^2}{16\pi^2m}\int^{k^{+}_F}_{k^{-}_F}dS_k\frac{k_x\frac{\partial\theta}{\partial
k_y}}{\Delta}\left(\mathbf{k}\times\frac{\partial\theta}{\partial\mathbf{k}}\right)_z.
\end{equation}
Comparison of Eq. (\ref{hyperLcon}) and Eq. (\ref{hyperScon})
gives $\sigma^{\mathscr{S}_z}_{xy}=-\sigma^{L_z}_{xy}$. As a
result, we have
$\sigma^{I_z}_{xy}=\sigma^{\mathscr{S}_z}_{xy}+\sigma^{L_z}_{xy}=0$,
i.e., the hyper-angular momentum current is zero.



\end{document}